\documentclass[aps,prb,twocolumn,groupedaddress,showpacs]{revtex4}
\usepackage{epsfig}
\usepackage{graphicx}% Include figure files
\usepackage{dcolumn}% Align table columns on decimal point
\usepackage{bm}% bold math
\usepackage{appendix}
\newcommand{\ii}{\'{\i}}

\newcommand{\vk}{{\bf k}}

\renewcommand{\vr}{{\bf r}}

\begin{document}
\title{Electron transmission in normal/heavy-fermion superconductor junctions}
\author{M. A. N. Ara\'ujo$^{1,2}$
  and P. D. Sacramento$^{1,3}$}
\affiliation{$^1$ CFIF, Instituto Superior 
T\'ecnico, TU Lisbon, Av. Rovisco Pais, 1049-001 Lisboa, Portugal}
\affiliation{$^2$ Departamento de F\'{\i}sica,  Universidade de \'Evora, P-7000-671, \'Evora, Portugal}
\affiliation{$^3$Departamento de F\ii sica, Instituto Superior 
T\'ecnico, TU Lisbon, Av. Rovisco Pais, 1049-001 Lisboa, Portugal}

\begin{abstract}
The Andreev reflection between a normal metal (N) and a heavy-fermion superconductor (HFS)
is studied and 
the boundary conditions  for the electron's wave function in the two systems
are established  in the framework of a two band model for the HFS. 
Hence we show in a simple and explicit way that the mass enhancement factors in the heavy-fermion (HF) metal do not cause impedance at the N/HFS interface, in accordance with arguments previously presented.
We also present an extension of the theory to a two-fluid model for the heavy-fermion, as 
possibly applicable to {\it e.g.}, CeCoIn$_5$.
\end{abstract}

\maketitle

\section{Introduction and model}

Electronic scattering at the interface between a normal metal (N)  and a superconductor has been studied  by Blonder, Tinkham and  Klapwijk\cite{blonder1982} (BTK) and it was found that
the subgap conductance is enhanced due to Andreev reflection.
On the other hand, the Fermi velocity mismatch between the two metals always produces an
effective barrier which decreases the conductance \cite{blonder1982,araujo2007}.  
If the superconductor is a heavy-fermion (HFS), the greater Fermi velocity mismatch 
 would lead us to expect a strongly reduced  subgap conductance. Experimentally, however, the subgap conductance does not seem to be strongly reduced in N/HFS junctions.
An  argument to explain this behavior has been put forward by Deutcher and 
Nozi\'eres\cite{nozieres}, who claimed that mass enhancement factors in the heavy-fermion 
metal do not cause impedance at the interface.

Motivated by recent experiments\cite{park2005,park2007} on Au/CeCoIn$_5$
interfaces, we study electron scattering at the interface between a normal (light) metal (N)  and a heavy-fermion superconductor. Starting from  a more realistic two-band model for the HFS, where a conduction  $c$-electron band hybridizes with  a localized $f$-electron band, plane-wave solutions for  each bulk subsystem can be written down.  
We explicitly obtain the matching conditions for the wave function at the interface and confirm the claim in Ref\cite{nozieres}, by explicitly showing that the  electron velocities involved are those of the conduction $c$-bands.  Since the localized $f$-electrons are dispersionless, only the $c$-conduction electrons satisfy matching conditions at the interface. This is the basis of the result.

An extension to a two-fluid model of the HFS, where a normal light fluid coexists with the superconducting one, is also presented, which may be relevant to {\it e.g.}, CeCoIn$_5$.
Our theory is  analogous to the  quantum waveguide theory for mesoscopic structures\cite{xia1992}.

The  model for the HFS is:
\begin{eqnarray}
\hat H_{hf} & = & \sum_{\vk\sigma} \left(\epsilon_\vk - \mu\right) 
c_{\vk \sigma}^{\dagger} c_{\vk\sigma}
+ \sum_{\vk\sigma} (\epsilon_f - \mu) f_{\vk \sigma}^{\dagger} f_{\vk\sigma}  \nonumber \\
&+& V \sum_{\vk\sigma} \left( f_{\vk \sigma}^{\dagger} c_{\vk\sigma} + h.c. \right) 
\nonumber \\ & + &  
\sum_{k} \left( \Delta_\vk f_{\vk \uparrow}^{\dagger} 
f_{-\vk,\downarrow}^{\dagger} + h.c. \right)\,.
\label{Hhfmodel}
\end{eqnarray}
The superconducting pairing, $ \Delta_\vk$,  has been explicitly written for the $f$-electrons  because such a model qualitatively reproduces the generic phase diagram of 
a HFS\cite{urbano2007,araujo2002,sacramento2003}. 
Equation (\ref{Hhfmodel})  is an effective model obtained in the large Anderson-$U$ limit, where it is understood that both the $f$-electron level, $\epsilon_f$, and the hybridization, $V$, are obtained self-consistently.
 
The Bogolubov  operators which diagonalize the above Hamiltonian are given by:
\begin{equation}
\hat \gamma_{\vk\sigma} =  u \hat c_{\vk\sigma} -\sigma v^* \hat c_{-\vk-\sigma}^\dagger  +
\tilde u \hat f_{\vk\sigma} -\sigma \tilde v^* \hat f_{-\vk-\sigma}^\dagger  \,,
\label{gamas}
\end{equation}
where the amplitudes obey the linear system
\begin{equation}
{\bf H}(\vk)\left(\begin{array}{c} u \\ \tilde u  \\ v \\ \tilde v\end{array} \right)
= E(\vk)\left(\begin{array}{c} u \\ \tilde  u \\ v \\ \tilde v\end{array}\right)\,,
\label{sistem}
\end{equation}
with
\begin{equation}
{\bf H}(\vk)=\left(\begin{array}{cccc}
\epsilon_\vk - \mu &  V & 0 & 0 \\
V & \epsilon_f - \mu & 0 & \Delta_\vk \\
0&0& -\epsilon_\vk + \mu & -V\\
0&  \Delta_\vk^* & -V & - \epsilon_f + \mu 
\end{array}\right)\,.
\label{Hmatrix}
\end{equation}
The excitation energies are given by: 
\begin{equation}
E(\vk)= \sqrt{\gamma(\vk) \pm \sqrt{\gamma^2(\vk)- \beta^2(\vk)}}\,,
\label{bands}
\end{equation}
with
\begin{eqnarray}
2\gamma(\vk) & = & (\epsilon_f - \mu)^2 + (\epsilon_\vk - \mu)^2 + |\Delta_\vk|^2+
2 V^2\,, \label{gama}\\
\beta^2(\vk)  & = & (\epsilon_\vk - \mu)^2 |\Delta_\vk|^2+ \left[ 
(\epsilon_f - \mu)(\epsilon_\vk - \mu) - V^2\right]^2\,.\nonumber\\ \label{beta} 
\end{eqnarray}

For convenience, we introduce the simplified notation:
\begin{equation}
\xi=\epsilon_\vk - \mu\,, \qquad \epsilon= \epsilon_f - \mu\,,\qquad \Delta = \Delta_\vk\,.
\end{equation}
The linear system (\ref{sistem})-(\ref{Hmatrix}) yields a relation between the 
amplitudes on the $f$ and $c$ subsystems:
\begin{eqnarray}
\tilde u  =  \frac{E-\xi}{V} u\,, \qquad \tilde v  =  -\frac{E+\xi}{V} v\,.
\label{uvtil}
\end{eqnarray}
From (\ref{uvtil}) we see that the amplitudes on the $f$ sites are not
independent being proportional to the amplitudes on the $c$ sites. 
The description of the bands is standard and is briefly reviewed in the Appendix.

\section{Electron transmission}
\label{Electrontransmission}

An incident electron from the light metal, with energy $E\geq 0$
measured from the Fermi level,  will penetrate the HFS. 
The coherence factors  for the transmitted quasiparticles (quasi-hole and quasi-electron)  
can be obtained from the linear system (\ref{sistem})-(\ref{Hmatrix}). 
Using (\ref{uvtil}) to eliminate $\tilde u$, $\tilde v$ we get:
\begin{eqnarray}
\left[ V^2-(E-\xi)(E-\epsilon)  \right] u - \Delta\cdot (E+\xi) v &=& 0\nonumber\\
\Delta\cdot (\xi -E) u + \left[ V^2-(E+\xi)(E+\epsilon)  \right] v &=& 0
\label{uvtrans}
\end{eqnarray}
Equating the corresponding determinant to zero we obtain:
\begin{equation}
\xi = \frac{\epsilon V^2 \pm \sqrt{-\Delta^2V^4 + E^2 \left( \epsilon^2+ \Delta^2+ V^2-E^2\right)^2}}{\epsilon^2+ \Delta^2 - E^2}
\label{xidentro}
\end{equation}
Equation (\ref{xidentro}) determines the momenta  of the transmitted quasiparticles
 (equation (\ref{xpos}) below).

At the Fermi level ($E=0$), the quasi-particles decay exponentially  into the HFS if the argument of the square root is negative (roughly if $E<\Delta_k$, for a specific direction).
Equation (\ref{xidentro}) then gives
\begin{equation}
\xi = \frac{\epsilon V^2 \pm \sqrt{-\Delta^2V^4}}{\epsilon^2+ \Delta^2} 
\approx \frac{V^2}{\epsilon} \pm i  \frac{V^2}{\epsilon^2}\Delta \approx\xi_F 
\pm i\hbar v_F^{(c)}\kappa\,,
\label{xitrans}
\end{equation}
where we have defined
\begin{equation}
\kappa = \frac{\Delta}{\hbar v_F^{(c)}} \left( \frac{V}{\epsilon}\right)^2\,.
\label{kappa}
\end{equation}
It is seen from equation (\ref{kappa}) that the momentum on the decaying quasiparticle 
in the HFS has a real part which is 
$k_F$ and an imaginary part, $\kappa$. The decay length, $\kappa^{-1}$, 
is determined by the {\it slow} Fermi velocity of the heavy-fermion system , 
$ v_F^{(c)}\epsilon^2/V^2$.

We write the electron wave function in the N/HFS system as a four-component vector by generalizing the column vector in equation (\ref{sistem}):
$$
\Psi(\vr)  =   \left(\begin{array}{c} f(\vr)\\ \tilde f(\vr)\\ g(\vr)\\ \tilde g(\vr) \end{array}\right)\,.
$$
Inside the normal (light) metal we have $\tilde f=\tilde g=0$, of course. 
We identify the N/HFS interface with the $x=0$ plane. When a plane wave is 
scattered at the interface, the wavevector component that is parallel to the 
interface is conserved. 
While no boundary conditions are imposed on $\tilde f $ and $ \tilde g$, 
the functions $f$ and $g$ are required to be continuous at the NS boundary.
Allowing for two different  $c$-band effective masses in the two metals, we write the kinetic
energy operator as\cite{nozieres,morrow} 
\begin{eqnarray}
\hat T = -\frac{\hbar^2}{2} \left[\partial_x\frac{1}{m(x)}\partial x + \frac{1}{m(x)}\left(\partial_y^2 + \partial_z^2\right)\right]\,,
\label{kinetic}
\end{eqnarray} 
where $m(x)=m_n$ in the light metal ($x<0$) and  $m(x)=m_{hf}$ in the heavy metal ($x>0$). 
The mass $m_{hf}$ is not large: it is simply the $c$-band effective mass. Because of the above form
of the kinetic energy, the derivative of $\psi$ is not continuous at $x=0$. The functions
$f$, $\tilde f$  and $g$, $\tilde g$ are coupled by the equations:
\begin{eqnarray}
Ef &=& \left[\hat T  -\mu + U\delta(x)\right] f + V\Theta(x) \tilde f\label{interface1}\\ 
Eg &=& \left[-\hat T  +\mu - U\delta(x)\right] g - V\Theta(x) \tilde g\,,
\label{interface2}
\end{eqnarray}
where $U$ denotes a potential barrier at the interface and $\Theta(x)$ denotes Heaviside's function. The remaining 
boundary condition for $f$ and $g$ is obtained upon integration of (\ref{interface1})-(\ref{interface2})  between $x=0^-$ and $x=0^+$:
\begin{eqnarray}
\frac{\hbar^2}{2}\left[ \frac{f'(x=0^+)}{m_{hf}} - \frac{f'(x=0^-)}{m_n}\right] 
&=& U f(0)\label{match1}\\
\frac{\hbar^2}{2}\left[ \frac{g'(x=0^+)}{m_{hf}} - \frac{g'(x=0^-)}{m_n}\right] &=& U g(0)\,.
\label{match2}
\end{eqnarray}
where the prime denotes derivation with respect to $x$. 

We write $\Psi(\vr) = e^{i\vk_{||}\cdot \vr_{||}} \psi(x)$ where $\vk_{||}$ and $ \vr_{||}$ are the wavevector and position vector components parallel to the interface, and
\begin{eqnarray}
\psi(x<0) & = &  \left(\begin{array}{c} 1\\ 0\\ 0\\ 0 \end{array}\right) e^{ip^+x} +
b  \left(\begin{array}{c} 1\\ 0\\ 0\\ 0 \end{array}\right) e^{-ip^+x} +
a\left(\begin{array}{c} 0\\ 0 \\ 1\\ 0 \end{array}\right) e^{ip^-x}\,,\nonumber\\
\psi(x>0) & = &  c\left(\begin{array}{c} u_+\\ \tilde u_+\\ v_+\\ \tilde v_+ \end{array}\right) e^{ik^+x} +
d  \left(\begin{array}{c} u_-\\ \tilde u_-\\ v_-\\ \tilde v_- \end{array}\right) e^{-ik^-x}
\,,\label{xpos}
\end{eqnarray}
where $b$ denotes the particle-particle reflection amplitude and $a$ denotes the 
Andreev reflection amplitude. The transmitted quasi-electron and quasi-hole in the 
HFS have amplitudes $c$ and $d$, respectively. The momenta are obtained from energy conservation:
$E(k_{||}, k^-)= E(k_{||}, k^+) = -\xi_n(k_{||}, p^-) = \xi_n(k_{||}, p^+)=E$, where
$E(\vk)$ is given in equation (\ref{Ek}) and $\xi_n(\vk)$ denotes the incident electron 
dispersion in the normal metal measured from the Fermi level.

In the following we calculate the reflection amplitudes $a$ and $b$ for $E=0$ and 
normal incidence ($p_{||}=0$), for simplicity. 
In this case all functions depend on the coordinate $x$, only. 
Equations (\ref{uvtrans})-(\ref{xidentro})  give $v_{+}=-iu$ and $k^+=k_F + i\kappa$ for the transmitted quasi-electron, and
$v_{-}=iu$ and $k^- =k_F - i\kappa$ for the transmitted quasi-hole.
Owing to the different $c$ band effective masses, we denote the Fermi velocity in 
the light metal by $v_n$, which should be comparable to that obtained from the slope of the $c$ band in the HF, $v^{(c)}_F$.
The continuity of the functions $f$ and $g$ at the interface implies:
\begin{eqnarray}
1+b&=&c u + d u\,,\\
ia&=&cu-du\,,
\end{eqnarray}
and equations (\ref{match1})-(\ref{match2}) imply:
\begin{eqnarray}
\frac{2U\left(1+b\right)}{\hbar^2}+ \frac{ip_F\left(1-b\right)}{m_n}&=&
cu\frac{ik^+}{m_{hf}} -du\frac{ik^-}{m_{hf}}
\,,\\
\left(\frac{2U}{\hbar^2} + \frac{ip_F}{m_n}\right) a &=&
cu\frac{k^+}{m_{hf}} +du\frac{k^-}{m_{hf}}\,,
\end{eqnarray}
where $\hbar p_F$ denotes the Fermi momentum in the normal metal,  $v_n=\hbar p_F/m_n$
and $v^{(c)}_F=\hbar k_F/m_{hf}$. 
Introducing the dimensionless barrier parameter\cite{blonder1982}
\begin{equation}
Z=\frac{U}{\hbar v_n}\,,
\end{equation}
we may write the reflection amplitudes as:
\begin{eqnarray}
a= \frac{-2i \ \frac{v^{(c)}_F}{v_n}}
{1+ \left(\frac{v^{(c)}_F}{v_n} \right)^2 + 
\left(2Z+ \frac{v^{(c)}_F}{v_n}\frac{\kappa}{k_F}\right)^2}\,,\label{a}\\
b=
\frac{\left( 1-2iZ-i\frac{v^{(c)}_F}{v_n}\frac{\kappa}{k_F}\right)^2-
\left(\frac{v^{(c)}_F}{v_n}\right)^2}
{1+ \left(\frac{v^{(c)}_F}{v_n} \right)^2 + 
\left(2Z+ \frac{v^{(c)}_F}{v_n}\frac{\kappa}{k_F}\right)^2}\,.\label{b}
\end{eqnarray}
The expressions  (\ref{a}) and (\ref{b}) are precisely what would be obtained if the 
HFS was a one-band superconductor with a Fermi velocity $v^{(c)}_F$. 
This can be traced back to the fact that only the $c$-electron parts of the wave function, 
$f(\vr)$, and $g(\vr)$, satisfy matching conditions
at the interface that are the same as in the case of a one-band superconductor.
Inside the HFS, the $f$-site amplitudes $\tilde f(\vr)$ and $\tilde g(\vr)$
are directly proportional to the $c$-electron amplitudes as shown in (\ref{uvtil}).
The mass enhancement, or slow Fermi velocity, in the HF appears in  (\ref{a}) and (\ref{b}) through
the  coherence length $\kappa$, as can be seen from (\ref{kappa}). 
 In the case of a clean interface ($Z=0$) 
and long decay length (such as close to a nodal direction), $\kappa\rightarrow 0$, these expressions simplify to
\begin{eqnarray}
a =\frac{-2i \eta}{1+\eta^2 }\,, \qquad b= \frac{1-\eta^2}{1+\eta^2}\,,
\label{BTKlimit}
\end{eqnarray}
where $\eta=v^{(c)}_F/v_n$.
If there is no mismatch of light Fermi velocities ($\eta=1$) then $b=0$ and
$|a|^2=1$ leading in this rather special case to a perfect doubled conductance:
$1+|a|^2-|b|^2=2$.

\section{Two-fluid HFS}
\label{TwofluidHFS}

A theory for electron transmission from a normal metal to a one-band 
superconductor\cite{kashiwaya} was applied to an interface with the
HFS CeCoIn$_5$\cite{park2005,park2007}. This material has a complex
Fermi surface \cite{dHvA} and seems to be well described by a two-fluid
model \cite{tanatar,pines,park2007}. 
The superconducting gap has $d$-wave symmetry, as determined directly by various
experiments \cite{curro,xiao,park2007,qpt}.

The HF metal CeCoIn$_5$ is known to have at least one band of uncondensed light carriers
in addition to the heavy fermion superconducting liquid. Recent Andreev reflection studies
assume that the subgap  conductance is a weighted average between that given by the
BTK theory and a flat conductance due to the band of uncondensed 
fermions\cite{park2007}. 
This motivates us to analyze the case where the incident electrons from the 
normal metal  (labeled as system "1")  
can tunnel simultaneously to a c-band coupled to the 
$f$ subsystem (labeled "2") 
and to a light uncondensed conduction band (labeled "3"), 
coexisting in the heavy-fermion material.
Metal 1 is in the $x<0$ half-space and metals 2 and 3 are in the $x>0$ half-space, 
with the interface at $x=0$. 
The theory we employ is an extension of the
quantum waveguide theory of mesoscopic structures\cite{xia1992} where a delta
function potential is introduced at the interface and where one of the circuit branches,
"2", is the HFS model of equation (\ref{Hhfmodel}). 

The potential $U\delta(x-\varepsilon)$ is in metal 1 ($x<0$) and we shall take the limit 
$\varepsilon\rightarrow 0^-$.
The wave function in the normal single-band metal 1 has particle and hole 
components\cite{explic}:
\begin{equation}
\psi_1(x\leq\varepsilon)=\left(\begin{array}c 1\\0\end{array} \right)e^{ip^+x} + 
b\left(\begin{array}{c} 1\\0\end{array} \right)e^{-ip^+x} + 
a \left(\begin{array}{c} 0\\1\end{array} \right)e^{ip^-x}\,,
\end{equation}
and  
\begin{eqnarray}
\psi_1(x\geq\varepsilon)&=&\alpha\left(\begin{array}c 1\\0\end{array} \right)e^{ip^+x} + 
\beta\left(\begin{array}{c} 1\\0\end{array} \right)e^{-ip^+x}\nonumber\\ 
&+& \gamma\left(\begin{array}c 0\\1\end{array} \right)e^{ip^-x} + 
\delta\left(\begin{array}{c} 0\\1\end{array} \right)e^{-ip^-x}\,. 
\end{eqnarray}
The wave function in metal 2 is written in the same form as (\ref{xpos}):
\begin{eqnarray}
\psi_2(x\geq0) & = &  c\left(\begin{array}{c} u_{_+}\\ \tilde u_{_+}\\ v_{_+}\\ \tilde v_{_+} \end{array}\right) e^{ik^+x} +
d  \left(\begin{array}{c} u_{_-}\\ \tilde u_{_-}\\ v_{_-}\\ \tilde v_{_-} \end{array}\right) e^{-ik^-x}
\,,
\end{eqnarray}
and for the normal metal 3 we simply write a transmitted electron and hole:
\begin{equation}
\psi_3(x\geq 0)=t\left(\begin{array}{c} 1\\0\end{array} \right)e^{iq^+x} + 
t_a \left(\begin{array}{c} 0\\1\end{array} \right)e^{iq^-x}\,.
\end{equation}

The matching conditions at $x=\varepsilon<0$ give the equations:
\begin{equation}
\psi_1(\varepsilon^-)=\psi_1(\varepsilon^+)\,,
\end{equation}
\begin{equation}
-\frac{\hbar^2}{2m_n}\left[ \psi_1'(\varepsilon^+)- \psi_1'(\varepsilon^-)\right] 
+U \psi_1(\varepsilon) =0\,.
\end{equation}
Taking the limit $\varepsilon\rightarrow 0^-$ we obtain:
\begin{eqnarray}
\alpha+\beta &=& 1+b\,,\label{a+b}\\
\alpha-\beta &=& \frac{2m_nU}{i\hbar^2p^+} (1+b) +1-b\,,\\
\gamma +\delta &=& a\,,\\
\gamma -\delta &=& \left( \frac{2m_nU}{i\hbar^2p^-} + 1 \right) a\label{g-d}
\,.
\end{eqnarray}
We now proceed with the matching condition at $x=0$ using the theory in Ref\cite{xia1992}.
The single-valuedness of the wave function  at $x=0$ implies that:
\begin{eqnarray}
\alpha+\beta &=& c u_{_+} + du_{_-} \,,\\
\alpha+\beta &=& t\,,\\
\gamma +\delta &=& cv_{_+} + dv_{_-} \,,\\
\gamma +\delta &=& t_a
\,,
\end{eqnarray}
and the (probability) current conservation implies that:
\begin{eqnarray}
\frac{ip^+}{m_n}\left(\alpha-\beta\right) &=& \frac{ik^+}{m_{hf}}c u_{_+} -
\frac{ik^-}{m_{hf}} du_{_-} + \frac{iq^+}{m}t\,, \\
\frac{ip^-}{m_n}\left(\gamma -\delta \right) &=& - \frac{ik^+}{m_{hf}}cv_{_+} 
-\frac{ik^-}{m_{hf}} dv_{_-} -\frac{iq^-}{m}t_a
\,, 
\end{eqnarray}
where $m$ denotes the electron's effective mass in metal 3.
Equations (\ref{a+b})-(\ref{g-d}) allow the elimination of the amplitudes 
$\alpha, \beta, \gamma, \delta$. 

These equations can be applied again to the case of normal incidence at $E=0$. Defining the Fermi momentum and velocity in metal 3  as $\hbar q_F$ and $v_3=\hbar q_F /m_3$,           respectively, 
we obtain  modified results for the amplitudes $a$ and $b$:
\begin{eqnarray}a=
\frac{-2i \ \frac{v^{(c)}_F}{v_n}}
{1+ \left(\frac{v^{(c)}_F}{v_n} \right)^2 + 
\left(2Z+ \frac{v^{(c)}_F}{v_n}\frac{\kappa}{k_F}\right)^2
+\frac{v_3}{v_n} \left(\frac{v_3}{v_n} +2\right)}\nonumber\\
\label{newa}
\end{eqnarray}
\begin{eqnarray}b=
\frac{\left( 1-2iZ-i\frac{v^{(c)}_F}{v_n}\frac{\kappa}{k_F}\right)^2-
\left(\frac{v^{(c)}_F}{v_n}\right)^2-\left(\frac{v_3}{v_n}\right)^2}
{1+ \left(\frac{v^{(c)}_F}{v_n} \right)^2 + 
\left(2Z+ \frac{v^{(c)}_F}{v_n}\frac{\kappa}{k_F}\right)^2
+\frac{v_3}{v_n} \left(\frac{v_3}{v_n} +2\right)}\nonumber\\ 
\label{newb}
\end{eqnarray}
In the limiting case of  a clean junction with the normal incidence close to the nodal direction, $Z,\kappa\rightarrow 0$, the above expressions simplify to
\begin{eqnarray}
a =\frac{-2i \eta}{1+\eta^2 +\eta'\left(\eta'+2\right)}\,,\nonumber\\ 
\qquad b= \frac{1-\eta^2-\eta'^2}{1+\eta^2+\eta'\left(\eta'+2\right)}\,,
\label{newBTKlimit}
\end{eqnarray}
where $\eta=v^{(c)}_F/v_n$, as in equation (\ref{BTKlimit}), and $\eta'=v_3/v_n$.
\begin{figure}
\centerline{\includegraphics[width=8.0cm]{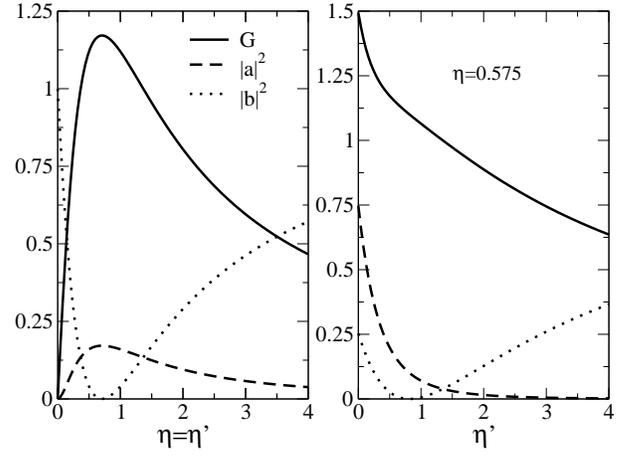}}
\caption{\label{cond2} 
Scattering amplitudes (\ref{newBTKlimit}) and conductance $G=1+|a|^2-|b|^2$ for
$Z,\kappa=0$ as
functions of $\eta'$ for:  $\eta=\eta'$ (left panel);  $\eta=0.575$ (right panel),
which is the value inferred from the effective barrier parameter estimated in Ref. \cite{park2007}.
}
\end{figure}
In order to obtain the perfect conductance for this case,
we assume no mismatch of light Fermi velocities ($\eta=\eta'=1$) so that $|b|^2=0.04$ and
$|a|^2=0.16$. The perfect conductance in this case is then 
\begin{equation}
G_{\it perfect}=1+|a|^2-|b|^2=1.12\,,\label{newconductance}
\end{equation}
which should be compared to the well known result $1+|a|^2-|b|^2=2$ in BTK theory. 
The effect of metal 3 is to "short-circuit" the system by providing a new 
transmission channel for the incoming electron. In this case the electron does not need to pick up  a second electron and penetrate the superconductor as a Cooper pair, leaving the Andreev hole behind. Hence
the  suppression of  the Andreev reflection process from $|a|^2=1$ to $|a|^2=4/25$. The effect of metal 3 is reduced if the ratio $v_3/v_n$ decreases and the theory of 
section \ref{Electrontransmission} is recovered in the limit $v_3/v_n \rightarrow 0$.
In the oposite limit, $v_3/v_n \rightarrow \infty$ implies $a\rightarrow 0$ and 
 $b\rightarrow -1$. Figure \ref{cond2} shows a plot of equations (\ref{newBTKlimit}) and
the conductance for some choices of velocity ratios.

We note that the result (\ref{newconductance}) is remarkably close to the enhanced
subgap conductance observed in CeCoIn$_5$, where the zero-bias differential 
conductance\cite{park2005, park2007} lies in the range 1-1.13. Figure \ref{cond2} shows
that some choices of velocity ratios yield conductances in this range. 
A fit  of our model  to the extensive experimental data on this compound might settle the model parameters values.  
 
We also note that  the form of equations 
(\ref{newa})-(\ref{newb}) shows that the conductance cannot be written a sum of two 
independent parallel conductances, one from the heavy liquid, another from the light one. The interface impedance here is a quantum-mechanical effect due to the boundary conditions for the wave function at an intersection, as in quantum waveguide theory.     

\section{Conclusions}

The success of the two-fluid model used in Ref. \cite{park2007}
to explain the conductance spectra of the junction Au/CeCoIn$_5$ relies 
on the assumptions that:
({\it i}) only the light velocities are important; ({\it ii}) the density of states is decreasing as one approaches the Fermi level;
({\it iii}) the conductance may be obtained as a weighted average.
Here we have explicitly confirmed the validity of  ({\it i}).
Item ({\it ii}) is not consistent with the usual approach where one considers the lowest
band partially filled, but is consistent if we consider that the lowest
band is full and the Fermi level is located at the bottom of the higher
band. In this regime the effective mass is also high and the density of
states is decreasing with energy. However, only the light velocities
affect the Andreev reflection.
Regarding ({\it iii}), we have shown that in general the conductance may not be written
as a sum since the interface impedance is a quantum-mechanical effect which
leads to interference terms.

\section{Acknowledgments}

We would like to thank Antonio C. Neto for bringing our attention to this problem
and V\'{\i}tor R. Vieira for discussions. 
This work was supported by Funda\c{c}\~ao para a Ci\^encia e Tecnologia 
(grant PTDC/FIS/70843/2006).

\appendix
\label{apa}
\section{Analysis of Eq. (\ref{bands})}
\begin{figure}[htb]
\centerline{\includegraphics[width=9.0cm]{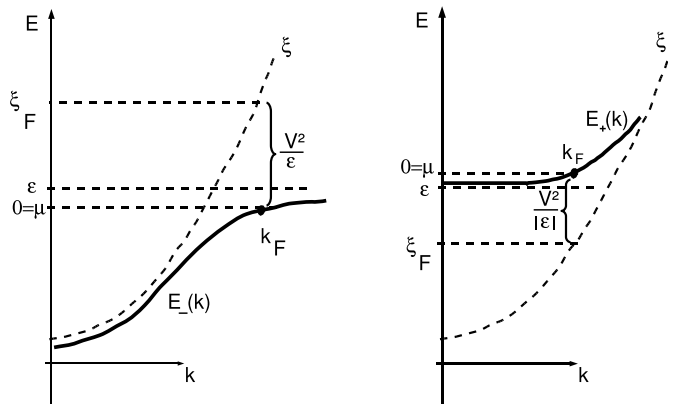}}
\caption{\label{HFbandas} 
Left: the lower partially occupied band $E_-(\vk)$ of a normal HF system resulting from 
the hybridization between the $f$ and $c$ subsystems with energies 
$\epsilon\equiv \epsilon_f - \mu$ and $\xi\equiv\epsilon_\vk - \mu$, respectively. 
The $c$ band energy evaluated at the Fermi momentum, $k_F$, is $\xi_F$ and has 
a slope $v_F^{(c)} = (d\xi/dk)_F$ which is larger
than the actual Fermi velocity $ \hbar^{-1}(dE_- /dk)_F= v_F^{(c)}\epsilon^2/V^2$.
Right: the upper partially occupied band $E_+(\vk)$.
}
\end{figure}

If $\Delta=0$ and the system is less than half-full,
all electrons are in the lower band given by
$$
E_- (\vk)= \frac{1}{2}\left[ \xi+ \epsilon -\sqrt{(\xi-\epsilon)^2 + 4V^2}\right]
$$
The Fermi level is given by
$E_- =0$, which means that $\beta$=0 in equations (\ref{bands}) and (\ref{beta}), or:
\begin{equation}
\xi_F=\frac{V^2}{\epsilon}
\label{xif}
\end{equation}
The Fermi velocity is then given by
\begin{equation}
\label{fermiv}
\frac{dE_-}{dk}=\left(\frac{dE_-}{d\xi}\right)_F \left(\frac{d\xi}{dk}\right)_F = 
\frac{1}{1+\left(\frac{V}{\epsilon}\right)^2}\hbar v_F^{(c)}
\end{equation}
where $ v_F^{(c)}=\hbar^{-1}\left( d\xi /dk\right)_F$ denotes the $c$-band  velocity evaluated at the  Fermi momentum $\hbar k_F$, and is comparable to that of the light metal. In order for the density of states (or mass enhancement) to be high, we must have  $V^2 \gg \epsilon^2$. Figure \ref{HFbandas} shows the bands in the normal state of the HF and the relevant energy scales. 

For finite $\Delta$, the lowest band of excitation energies is obtained by choosing the minus sign in equation (\ref{bands}), which may be written as
\begin{equation}
E_-^2 =\gamma_0 + \frac{\Delta^2}{2} -\sqrt{ \left(\gamma_0 + \frac{\Delta^2}{2}\right)^2 - 
\left( \beta_0^2 + \Delta^2\xi^2\right)}
\end{equation}
where $\gamma_0$ and $\beta_0$ are the functions $\gamma$ and $\beta$ given in
(\ref{gama})-(\ref{beta})  when $\Delta=0$. By expanding in $\Delta$ (assuming 
$V^2 \gg \epsilon^2 \gg \Delta^2$ which is valid for instance for CeCoIn$_5$ where
the coherence temperature is of the order of $45$ K and the superconducting critical
temperature is of the order of $2.3$ K) we obtain the gap for excitations at $k_F$:
\begin{equation}
E(\vk_F)= \Delta(\vk_F) \frac{\left(V/\epsilon\right)^2}{1+\left(\frac{V}{\epsilon}\right)^2} \approx \Delta(\vk_F)
\end{equation}
and the excitation spectrum in the vicinity of the Fermi level is:
\begin{equation}
E(\vk)= \sqrt{\left[\frac{\hbar v_F^{(c)}\left( |\vk| - k_F\right)}{1+\left(\frac{V}{\epsilon}\right)^2}\right]^2 + \Delta^2(\vk)} \,,
\label{Ek}
\end{equation}
which has the usual form and involves the slow velocity of the heavy electrons.

We may as well consider a situation where the local Coulomb repulsion between
the electrons is not too large, which enables a $f$-site occupancy $n_f>1$. This case is
also represented in Fig. \ref{HFbandas}, where we consider that the Fermi level is located above  $\epsilon_f$. In this case $\xi_F,\epsilon<0$. The position
of the Fermi level is again obtained from $\beta_0=0$ and leads in this case to
 $\xi_F=-V^2/|\epsilon|$. It is easy to see that the Fermi
velocity is once again of the form  (\ref{fermiv}), as expected of a heavy band. Note however,
that while for the case when $n_f<1$ the density of states is increasing as we approach
the Fermi level, in the case when $n_f>1$ the density of states is a decreasing
function of the energy in the vicinity of the Fermi level.

%%%%%%%%%%%%%%%%%%%
%    bibliography  
%%%%%%%%%%%%%%%%%%%%%%%%%%%%%%%%%%%%%%%%%%%%%%%%%%%%%%%%%%%%%%%%%%%%%%%%%%%%%%%


\begin{thebibliography}{99}
\bibitem{explic} The components $\tilde f$, $\tilde g=0$ and have, therefore, been omitted.
\end{thebibliography}
\end{document}